\documentclass[aps,physrev,reprint,superscriptaddress]{revtex4-2}

\usepackage{physics}
\usepackage{mathtools}
\usepackage{hyperref}
\usepackage{url}

\begin{document}

%\preprint{}

\title{A Quantum Key Distribution System for Mobile Platforms with Highly Indistinguishable States}



\author{Daniel Sanchez Rosales}
\affiliation{The Ohio State University, Department of Physics, 191 W. Woodruff Ave., Columbus, OH 43210, USA}

\author{Roderick D. Cochran}
\affiliation{The Ohio State University, Department of Physics, 191 W. Woodruff Ave., Columbus, OH 43210, USA}

\author{Samantha D. Isaac}
\affiliation{University of Illinois Urbana-Champaign, Department of Physics, 1110 West Green St., Urbana, IL 61801, USA}

\author{Paul G. Kwiat}
\affiliation{University of Illinois Urbana-Champaign, Department of Physics, 1110 West Green St., Urbana, IL 61801, USA}

\author{Daniel J. Gauthier}
\affiliation{The Ohio State University, Department of Physics, 191 W. Woodruff Ave., Columbus, OH 43210, USA}
\email[]{gauthier.51@osu.edu}

%\author{}
%\thanks{}
%\affiliation{}

\date{\today}

\begin{abstract}
Quantum key distribution (QKD) allows two users to exchange a provably secure key for cryptographic applications. In prepare-and-measure QKD protocols, the states must be indistinguishable to prevent information leakage to an eavesdropper performing a side-channel attack. Here, we measure the indistinguishability of quantum states in a prepare-and-measure three-state BB84 polarization-based decoy state protocol using resonant-cavity light-emitting diodes (RC-LEDs) as the source in the transmitter. We make the spatial, spectral, and temporal DOF of the generated quantum states nearly indistinguishable using a spatial filter single-mode fiber, a narrow-band spectral filter, and adjustable timing of the electrical pulses driving the RC-LEDs, respectively. The sources have fully indistinguishable transverse spatial modes. The measured fractional mutual information between an assumed eavesdropper and the legitimate receiver is $2.39\times10^{-5}$ due to the spectral distinguishability and $4.31\times10^{-5}$ for the temporal distinguishability. The source is integrated into a full QKD system operating in a laboratory environment, where we achieve a raw key rate of 532 Kbits/s with an average quantum bit error rate of 1.83\%. The low system size, weight, and power make it suitable for mobile platforms such as uncrewed aerial vehicles (drones) or automobiles.
\end{abstract}


\maketitle

\section{Introduction} \label{sec:intro}
Ensuring secure communication is crucial in our ever-growing, interconnected world, yet it remains consistently threatened by eavesdroppers who exploit system vulnerabilities to design novel attacks on our communications. The advent of quantum computers further compounds this issue because it is predicted that they can be used to compromise existing public-key cryptographic methods. Quantum Key Distribution (QKD) is one possible solution to this dilemma.  It is based on physics principles for securely sharing a random bit string between two users, which can be used as a cryptographic key for encrypting and decrypting plain-text messages.

\begin{figure}[htbp]
\centering
\includegraphics[width=0.8\linewidth]{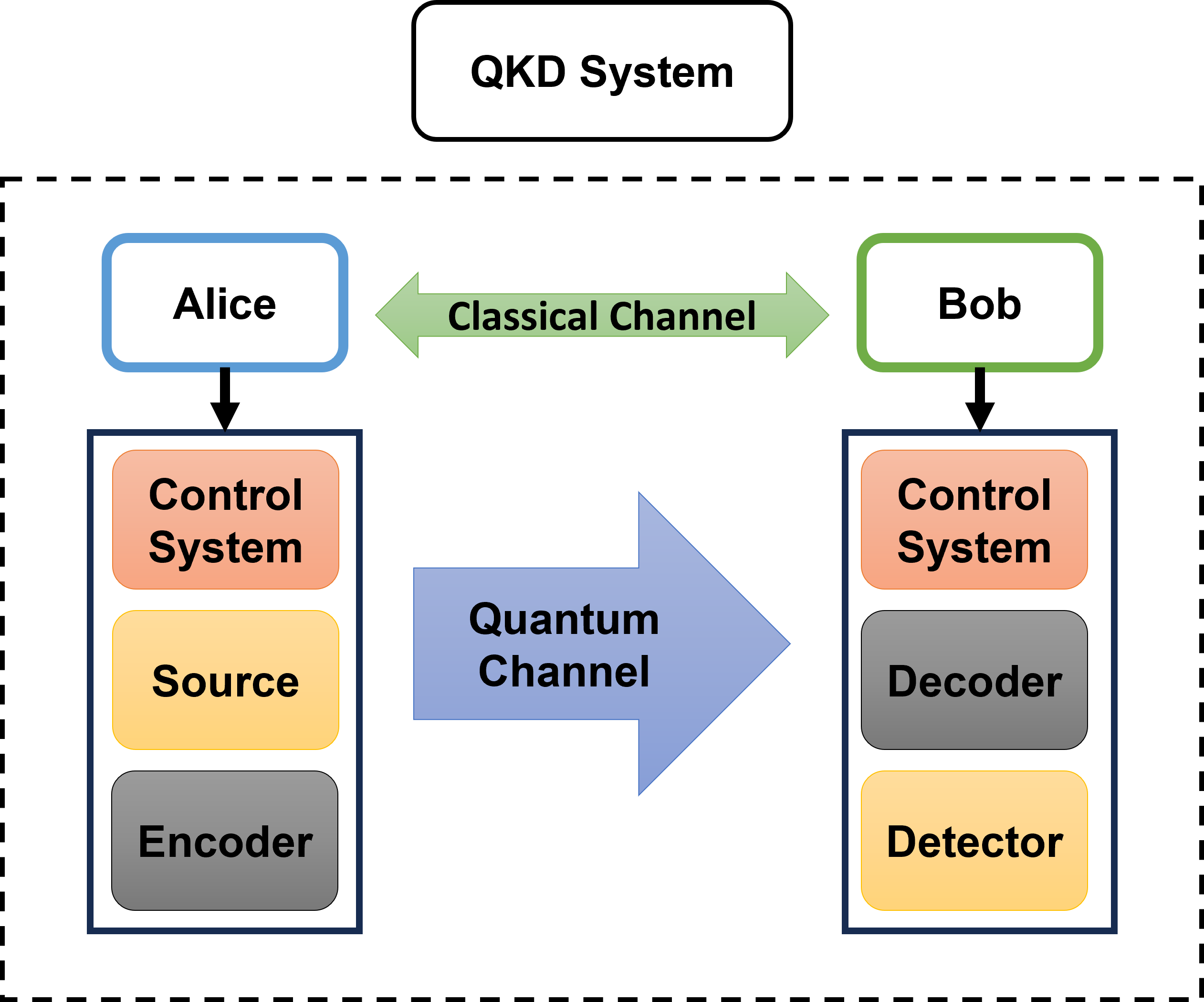}
\caption{A conceptual layout for a QKD system. Alice sends Bob quantum information through a quantum channel, and they both communicate through a classical, insecure, channel} 
\label{fig:qkd_big_picture}
\end{figure}

Currently deployed QKD systems allow point-to-point communication between a sender, commonly referred to as Alice, and a receiver, referred to as Bob.  As illustrated in Fig.~\ref{fig:qkd_big_picture}, Alice's transmitter consists of a classical control system, a source of quantum light, and a quantum state encoder. Bob's receiver consists of a control system, a quantum state decoder, and single-photon-counting detectors. Alice encodes a random bit sequence on quantum photonic states and sends them through a free-space or fiber quantum channel to Bob, who decodes and detects the states.  The transmitter and receiver are connected through a classical communication channel for control-system clock synchronization, reconciliation, error correction, and privacy amplification.

QKD systems are maturing rapidly following this general approach \cite{QKD_review}, including recent commercial systems \cite{toshiba, idquantique, qctek}. These studies and systems often focus on long-distance key distribution \cite{Liao2017, TFQKD, TimeBinQKD}, which will likely form the backbone of a future communication system with fixed and secure physical infrastructure.  The size, weight, and power (SWaP) of these systems are not pressing design metrics.

Yet, like classical communication systems, different solutions are likely required for cryptographic key exchange over the `last mile' between the closest backbone station and the user.  Some users will be mobile so the network needs to be reconfigurable to route the quantum states between Alice and Bob.  There is substantial progress on developing low-SWaP QKD systems for space \cite{Liao2017} and flight applications \cite{Pugh_2017}.  Also, prototype devices aimed at hand-held applications are under development \cite{Lowndes2021, Chun:17, Hand_Held, Xia}, as well as devices that can operate on small (Group 1 \& 2) uncrewed aerial systems typically referred to as drones \cite{Drone_Groups, drone_to_ground_entanglement, drone_to_ground_qkd}. 

Developing a low-SWaP system requires a full system analysis of the trade-off between system performance and SWaP. Some recent studies have relaxed system security requirements to achieve SWaP metrics with the idea that these attack vectors will be addressed in the future \cite{Lowndes2021, Chun:17, Hand_Held, Xia}.  Here, we present a system-level analysis, design, and performance characteristics of a QKD system intended for mobile applications.  We address some of the security issues associated with previous low-SWaP designs by making the quantum states generated by the transmitter highly indistinguishable, thus removing several side-channel attack vectors.  Our intended application is QKD between drones or automobiles, which is described elsewhere \cite{SPIE_1, SPIE_2}.

In the first part of this paper, we present the design considerations for the QKD system, such as choosing the QKD protocol, control system, and detectors. Our goal in this section is to provide a guide for the type of questions and challenges that arise when designing such a platform. We then discuss our experimental setup to demonstrate how the selected components are assembled and managed, and how data is generated, collected, and processed. Finally, we present laboratory bench results of a full QKD system and discuss future improvements to our setup.

\section{Design Considerations} \label{sec:design_considerations}
Here, we discuss the high-level design considerations in developing our QKD system, mentioning a few alternative solutions and discussing our final design choices. To the extent possible, we restrict our attention to commercially available components to shorten the design cycle.

\subsection{QKD Protocol} \label{sub:QKD_protocol}
Likely the most important metric for a QKD system is the secure key rate, which requires near-ideal quantum state preparation and measurement, no leakage of information into other degrees-of-freedom (DOF) of the quantum photonic wavepackets, and usually requires a high transmitter and receiver clock rate. Prepare-and-measure (P\&M) \cite{PMQKD_easy} and entangled-state protocols \cite{EB_QKD} are the most widely used QKD schemes, where entangled-state systems are more resilient to noisy channels but typically have lower key rates than P\&M QKD.  We select the P\&M protocol for its higher rate, which will reduce the connection time of the mobile platforms.  Furthermore, we focus on free-space optical channels because they are best suited to mobile users.

The most widely used P\&M QKD scheme is a cryptographic protocol developed by Bennett and Brassard in 1984, known as BB84 \cite{BB84}. The protocol relies on transmitting quantum photonic wavepackets in which the information is encoded in one or more of the photon DOF.  The most common approaches use the polarization or time-frequency states, where the latter requires using delay interferometers \cite{TaimurTimeBin}. These interferometers are often bulky and require high-precision optics, making them less suitable for low-SWaP systems.  Furthermore, we desire multi-mode receivers to increase the photon collection efficiency, but multimode unbalanced interferometers are not yet commercially available \cite{Clinton_interferometer}. 

Polarization-based protocols eliminate the need for interferometers and multi-mode quantum state decoders are readily available to enhance light collection efficiency. Thus, a polarization-based protocol is more suited for applications requiring high mobility and low SWaP. However, we must select the states to minimize errors due to the relative rotation of one platform relative to the other.

In many polarization-based applications, a single source passes through a polarization modulator to encode the quantum states. However, bulk modulators require high voltages, and have low bandwidth for devices working in the visible and near-infrared spectrum.  For this reason, we consider a transmitter with an independent source for each quantum state.

Multi-source P\&M protocols require indistinguishable sources to prevent side-channel attacks. In most cases \cite{Xia, Lowndes2021, Hand_Held, Chun:17}, this problem is acknowledged but not addressed or quantified. In this work, we make the states as indistinguishable as possible and quantify the associated information leakage to an eavesdropper, referred to as Eve, by estimating the mutual information between Eve and Bob.

Moreover, the classic BB84 protocol relies on assuming perfect single-photon sources.  Low-SWaP single-photon sources are not yet commercially available, so we focus on imperfect sources that sometimes produce multi-photon wavepackets.  For these wavepackets, Eve can perform a photon number splitting (PNS) attack \cite{PNS_attack}, obtaining potentially all information in these wavepackets. To mitigate this security risk, we use a decoy-state protocol developed to ensure secure communication with imperfect sources \cite{Decoy_Lo}.

In summary, we find that a free-space P\&M polarization-based BB84 scheme using multi-photon sources and the decoy state protocol is well suited for performing QKD on mobile platforms with low-SWaP restrictions. 

\subsection{Control Systems} \label{sub:control_systems}
We now discuss Alice and Bob's classical control systems.  There are several low-SWaP options for the control system, such as single-board computers, microcontrollers, and field-programmable gate arrays (FPGAs). To narrow our choice, we need to understand a few additional requirements of a QKD system.

One requirement for the control system for a multi-source QKD system operating above 10 MHz, typical for modern systems, is to generate multiple nanosecond-scale digital electrical pulses with sub-nanosecond relative timing accuracy that can be turned on in a random sequence.  Single-board computers and microcontrollers usually have an operating system that does not operate in real-time and hence has interrupts that can disrupt timing.  While some microcontrollers have independent timing/frequency units that are not disturbed by interrupts, it is difficult to have fine relative timing control across multiple channels.  This suggests using an FPGA, which can precisely adjust the timing of nanosecond electrical pulses and perform tasks in parallel.

Another advantage of using an FPGA is the ability to generate true random numbers at a high rate \cite{True_RNG}.  Here, the random number generator selects in real-time the symbol and basis to encode the quantum states and the selection of decoy states.  The random numbers must also be stored on the mobile platform because they form the raw key or are used in the security analysis.  In Sec.~\ref{subsub:RX_data_processing} below, we discuss how FPGAs tightly integrated with a hard processor (known as a system-on-a-chip) can achieve high-throughput data storage.

We use the Intel/Terasic DE10-Standard demonstration board for Alice and Bob's control systems \cite{DE10_web}; other system-on-a-chip devices will likely have similar performance characteristics.  The FPGA part of the chip has 5.6 Mbits of embedded memory, 6 dynamically adjustable phase-locked-loop (PPL) oscillators, and multiple input-output lines between the chip and standard board connectors. A high-speed bus connects the FPGA to the hard processor, a 925 MHz, Dual-Core ARM Cortex-A9 hard processor using a Linux operating system with 1 GB of random-access memory and a secure digital (SD) card for long-term storage. The demonstration board has dimensions 16.7 cm $\times$ 13.0 cm $\times$ 2.7 cm, weighs 0.19 kg, and, when in use during a QKD session, draws a power of 2.5 W (2.6 W) for Alice (Bob). 

\subsection{Detector} \label{sub:detector}

For discrete-variable quantum communication systems, we must record single-photon events at Bob's receiver.  We first discuss this design consideration because commercially available single-photon detectors operate over a limited spectral band and hence affect our source design discussed in the next section.  Many QKD systems, including some commercial systems, use superconducting nanowire single-photon detectors \cite{SNSPDs}, which have high detection efficiency and rate, and low noise.  Unfortunately, they require cryogenic temperatures ($\sim$1 K), thus greatly limiting their application for mobile QKD systems, although a drone-based SNSPD system was recently reported \cite{drone-SNSPD}.

We therefore focus on room-temperature single-photon avalanche photodiodes (SPADs).  SPADs are available that are sensitive to photons in the telecommunication band at a wavelength near 1.55 $\mu$m, but they tend to have high noise and low detection rates unless complex custom circuits are used to improve their performance \cite{SPADs_Shield}. In the visible part of the spectrum, silicon-based SPADs are commercially available, have high efficiency, and have a saturated detection rate greater than $\sim$5 MHz.  For these reasons, we select the Excilitas SPCM-AQ4C module with four independent detectors, which matches the polarization-based BB84 protocol that requires measuring four states.  They have a peak quantum efficiency at a wavelength of 650 nm and are available with pre-aligned multi-mode fiber couplers. It has dimensions of 17.8 cm $\times$ 14.1 cm $\times$ 3.4 cm, weighs 0.68 kg, and typically draws a power of 3.3 W, dominated by the thermoelectric cooler that lowers the temperature of the diode junction, thereby reducing dark counts.  

\subsection{Source} \label{sub:source}

We now focus on the source for our P\&M polarization-based protocol.  Several groups have reported using a laser for each state and basis (see, for example, Ref.~\cite{VCEL_FPGA, Xia, WCP_QKD}).  However, the laser spectrum tends to be narrow-band; adjusting them to be identical requires injection locking or precise spectral measurement, which may have to be repeated as the temperature changes.  

On the other hand, LEDs have a much broader spectrum that can then be made nearly indistinguishable using a narrow-band spectral filter.  We also spatially filter the light by coupling the LED emission into single-mode optical fibers, substantially reducing the source power.  We use resonant-cavity LEDs, which have a decreased emission angle and narrower spectrum to increase the spectral brightness and hence increase the single-spatial-mode power.  Some years ago, RC-LEDs were available commercially at a variety of wavelengths \cite{RCLEDs}, but we are aware of only one device available commercially, from Roithner LaserTechnik: the RC650-TO46FW.  This RC-LED operates at a wavelength of 650 nm, which conveniently matches the peak detection efficiency of the silicon SPADs. We use these RC-LEDs in our QKD transmitter. 

Wavepackets generated by a pulsed RC-LED attenuated to an average mean photon number of $\sim$1 required for QKD are described by thermal statistics; hence, some pulses will have more than 1 photon. The  QKD system can be secured even with these imperfect sources by using the decoy-state protocol \cite{Curty:09, Curty:10, WCP_laser}. Here, Alice randomly adjusts the mean photon number of the wavepackets. After a QKD session, Alice announces the mean photon number of each transmitted wavepacket.  Statistical analysis of Bob's received events allows them to place a tight bound on the fraction of wavepackets having more than one photon. Typically, a system uses a mean photon number $\sim$1 for the signal state, and two decoy states with mean photon numbers $\sim$0.4 and 0.  The probability of transmitting each state is non-uniform, where the signal state is most probable and the other state probabilities are adjusted to optimize the secure key rate.  From the analysis, Alice and Bob decide whether the measured statistics match their expected values for the mean photon number used in the experiment; a deviation indicates the presence of an eavesdropper.

Another advantage of using RC-LEDs is that they can be driven directly by the output of the FPGA, where the typical maximum current for the RC650-TO46FW is 20 mA.  Furthermore, the power emitted by the RC-LED is a linear function of the injection current, which simplifies generating the decoy states as described in Sec.~\ref{sub:FPGA_LED}.

\subsection{Encoder and Decoder} \label{sub:encoder_decoder}

To speed up system implementation, we use a small free-space optical bench consisting of a custom quarter-inch-thick aluminum plate onto which the encoder and decoder optics are placed.  On Alice's side, the encoder uses half-inch optics, a compromise between system size and weight while minimizing the diffraction of the light sent through the quantum channel.  Assuming an Airy pattern for the light emitted from the last optic, we estimate that we can achieve up to a 90\% collection efficiency at a receiver, also with half-inch optics, located a distance of $\sim$260 m from the source.  Longer distance QKD systems will require larger-diameter optics on the receiver or both the transmitter and receiver if diffraction is the primary loss mechanism. The encoder dimensions are 18.8 cm $\times$ 15.3 cm $\times$ 4.0 cm and weighs 1.7 kg when fully loaded with the encoder optics.

Bob's decoder uses one-inch optics to enhance collecting light from our pointing, acquisition, and tracking system (see Sec. \ref{sub:state_decoding} below). The decoder dimensions are 20.0 cm $\times$ 16.1 cm $\times$ 4.0 cm and weighs 1.8 kg fully loaded. 

\section{Implementation} \label{sec:experiemntal_setup}

\subsection{The Transmitter}

\begin{figure}[htbp]
\centering
\includegraphics[width=0.96\linewidth]{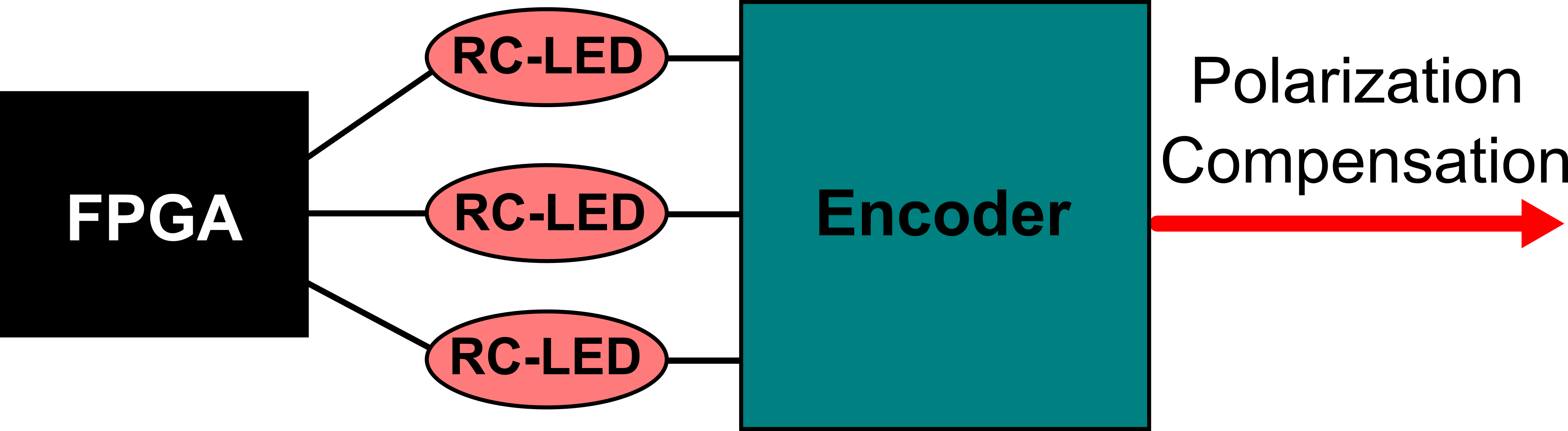}
\caption{High-level diagram showing the experimental setup for Alice's quantum transmitter.} 
\label{fig:high_level}
\end{figure}

Recent research has shown that sending only three quantum states in the BB84 protocol but measuring all four achieves the same secure key rate as sending four states \cite{Islam, Tamaki}. We adopt this protocol: our transmitter uses three separate RC-LEDs driven by the FPGA to produce optical pulses attenuated to the single-photon level. The RC-LEDs are coupled to single-mode fibers, which direct the light to the polarization-based free-space encoder (see Secs. \ref{subsub:LED_optical} and \ref{subsub:TX_optical}). After polarization encoding, the beams are combined along a common optical axis and coupled into a single-mode fiber to make the spatial modes of each state indistinguishable.  Finally, we compensate for the unitary polarization transformation induced by the spatial-filter fiber. These subsystems are illustrated in Fig.~\ref{fig:high_level} and discussed in the sub-sections below.

\subsubsection{FPGA Pulse Generation} \label{sub:FPGA_pulse_generation}

We drive the RC-LEDs using electrical pulses generated by dynamically reconfigurable PLLs on the FPGA \cite{Intel_PLL}.  These pulses have an amplitude of 3.3 V, a pulse width of 40 ns, and a repetition rate of 12.5 MHz (the QKD system clock frequency).  We use two PLLs, each with 6 outputs to drive the three RC-LEDs with the signal and non-zero decoy state. The delay of each PPL output can be adjusted with a resolution of 78 ps without recompiling the FPGA firmware. To adjust the pulse width, we combine two PLL outputs with an AND gate so that the relative delay of the resulting data pulse changes the width, where we use a typical pulse width of 10 ns. Small adjustments of the electrical pulses are made so that each state's quantum photonic wavepackets have essentially identical pulse widths as discussed in Sec.~\ref{sub:indistinguishability}.  We adjust the relative timing of one data pulse relative to the other by setting the delay of both PLL outputs used to generate a data pulse.

\begin{figure}[htbp]
\centering
\includegraphics[width=0.85\linewidth]{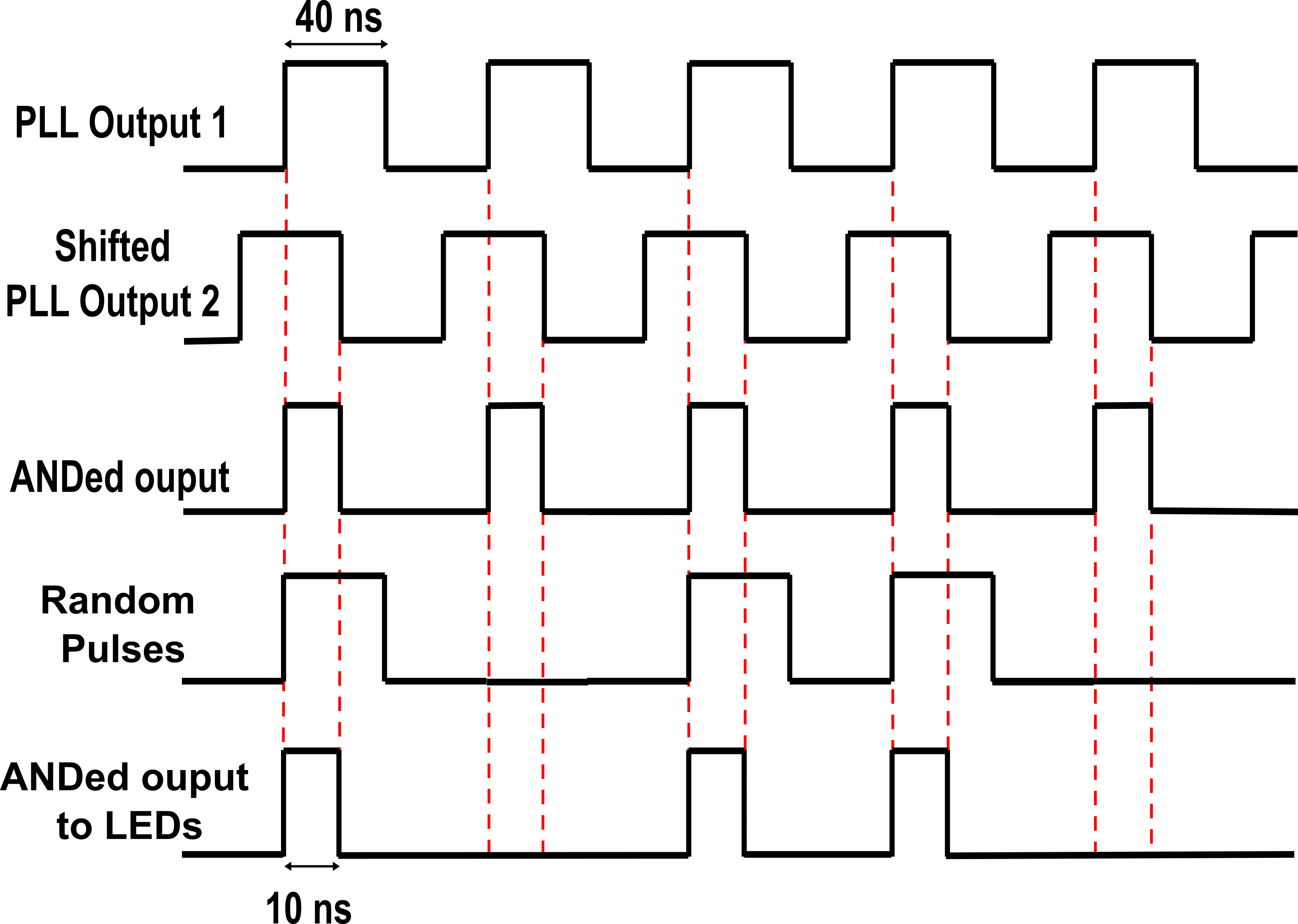}
\caption{Timing diagram illustrating the method used to generate adjustable and random electrical pulses that drive the RC-LEDs.} 
\label{fig:width_control}
\end{figure}

To select the transmitted state (one of the three polarization states, signal state, or decoy states), we use a digital true random number generator \cite{True_RNG} on the FPGA that produces a 40-ns-long pulse.  The pulse representing the randomly selected state and the data pulse are combined with an AND gate to unmask the desired state as illustrated in Fig.~\ref{fig:width_control}.

\subsubsection{FPGA-LED Interface} \label{sub:FPGA_LED}

The electrical pulses generated by the FPGA pass to the onboard general-purpose input-output (GPIO) connector pins and then to a custom circuit board, where three RC-LEDs are attached.  Each RC-LED is connected to two electrical transmission lines on the circuit board, each connected to separate GPIO pins.  This allows us to drive a single RC-LED with two different electrical waveforms. 

\begin{figure}[htbp]
\centering
\includegraphics[width=0.8\linewidth]{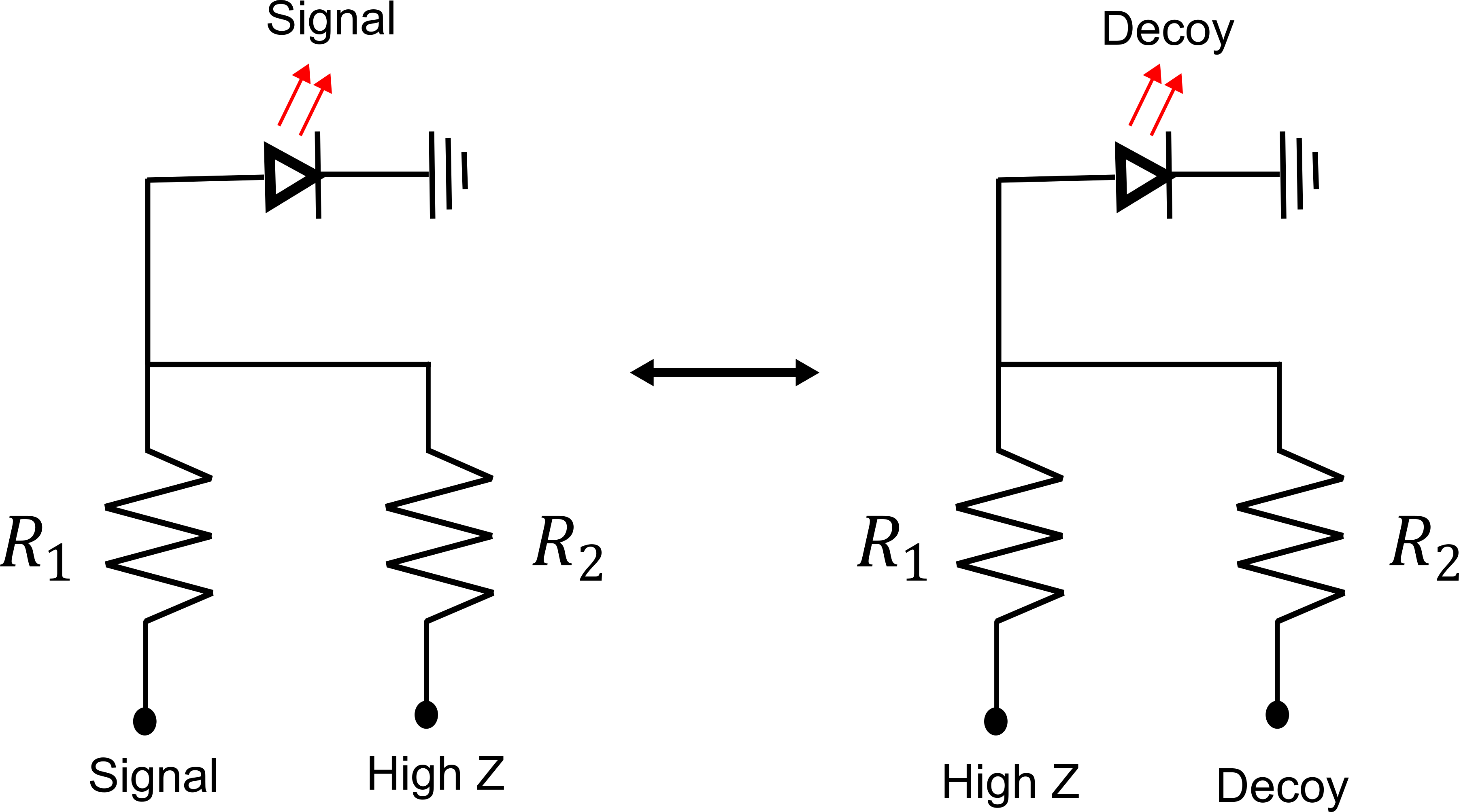}
\caption{Electrical circuit showing the different paths for the signal and decoy states.} 
\label{fig:LED_Resistive}
\end{figure}

As shown in Fig.~\ref{fig:LED_Resistive} for a single RC-LED, the pulse for the signal state passes through resistor $R_1$ and the RC-LED to ground.  The other path is set to a high impedance (high-$Z$) state so no current flows back to the FPGA through this path. Switching an output between the high- and low-impedance states occurs on the nanosecond timescale. To generate the decoy state, the signal path is set to the high-$Z$ state and the decoy-state pulse passes through resistor $R_2$ and the RC-LED to ground.  Here, $R_1$ = 20.5 $\Omega$ to produce an injection current of 25 mA, and $R_2$ = 182 $\Omega$ to produce an injection current of 10 mA so that the intensity of decoy states is $\sim$40\% lower than the signal states.  Using this architecture for each RC-LED, we produce 3 signal states, 3 decoy states, and 3 vacuum states using only 3 sources. 

The probability of selecting the signal, non-zero decoy, and vacuum decoy states. To this end, the true random number generator produces bits at a rate 8$\times$ higher than the QKD transmitter rate so we have 8 bits available each moment a state needs to be accepted.  

The 256 levels are used to select the state and the probability of generating the state. If the generated number is between 0 and 50 (inclusive, decimal representation), we choose the state signal $\ket{R}$, the state $\ket{L}$ when it is between 51 and 101, and the state $\ket{H}$ when it is between 102 and 179. Similarly, we send the state decoy $\ket{R}$ when the generated number is between 180 and 193, the state decoy $\ket{L}$ when it is between 194 and 207, and the state decoy $\ket{H}$ when it is between 208 and 228. Finally, we send the vacuum state when the generated number is between 229 and 255.

\subsubsection{LED-Optical Setup Interface} \label{subsub:LED_optical}

\begin{figure}[htbp]
\centering
\includegraphics[width=1.0\linewidth]{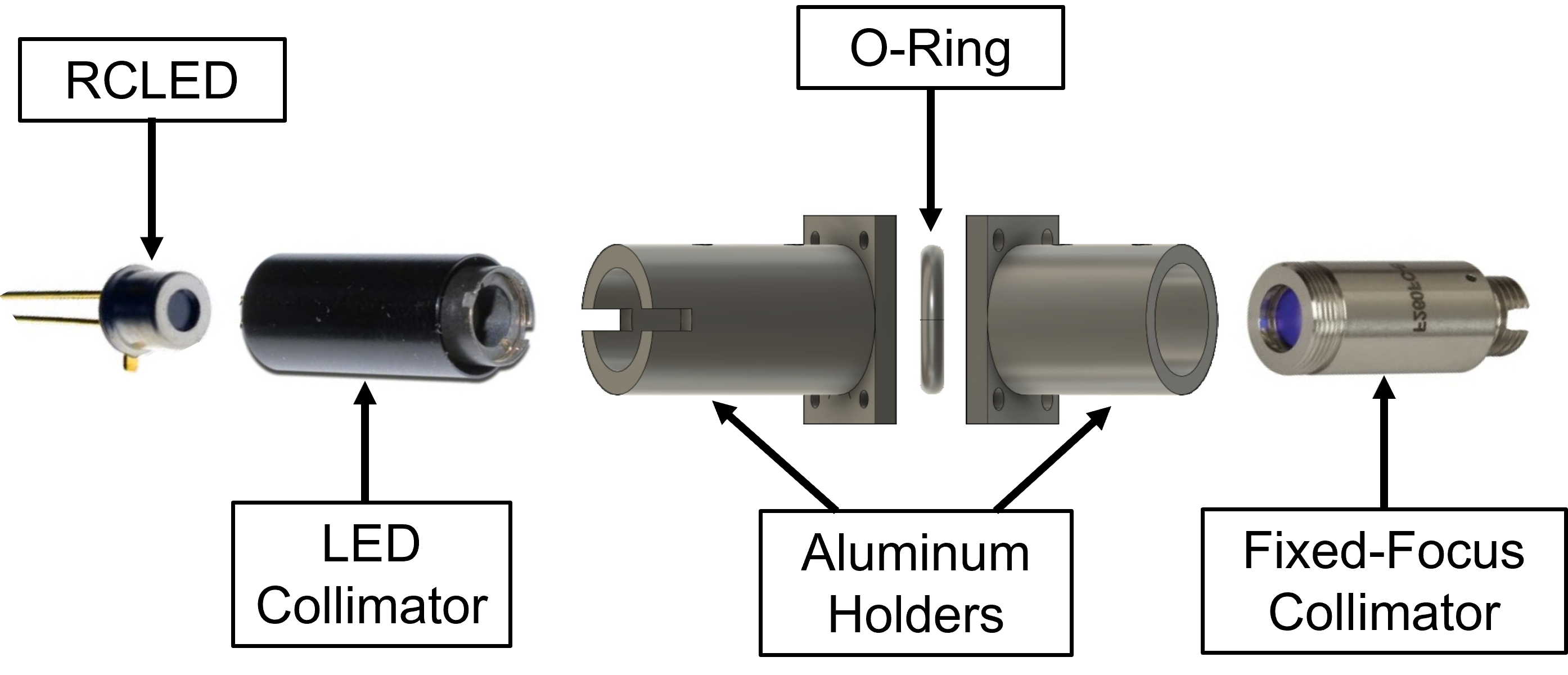}
\caption{Exploded diagram of the custom-built LED-SMF coupler.} 
\label{fig:LED_Coupler_Expanded}
\end{figure}

We separate the control system and RC-LED sources from the free-space optical bench to minimize the weight of the sub-system containing the quantum state encoder.  We use single-mode fibers (SMFs) to deliver the light from the source to the encoder. To improve the collection efficiency of light coupled into the fibers, we develop a coupler consisting of the RC-LED in its corresponding collimation package (RC-LED-650-02) and a fixed-focus collimation package (Thorlabs F220FC-B) to focus the light on the core of the SMF. We fit both of these collimators in a custom-built aluminum holder, which is spaced by an o-ring and attached with screws as illustrated in Fig.~\ref{fig:LED_Coupler_Expanded}.  Adjusting the screws allows us to maximize the coupling efficiency into the SMF.  We use an in-fiber variable attenuator (OZ Optics BB-700-11-650-4/125-S-40-3S3S-1-HYBK-0.25) to adjust the wavepacket to the single-photon level.

\begin{figure}[htbp]
\centering
\includegraphics[width=0.8\linewidth]{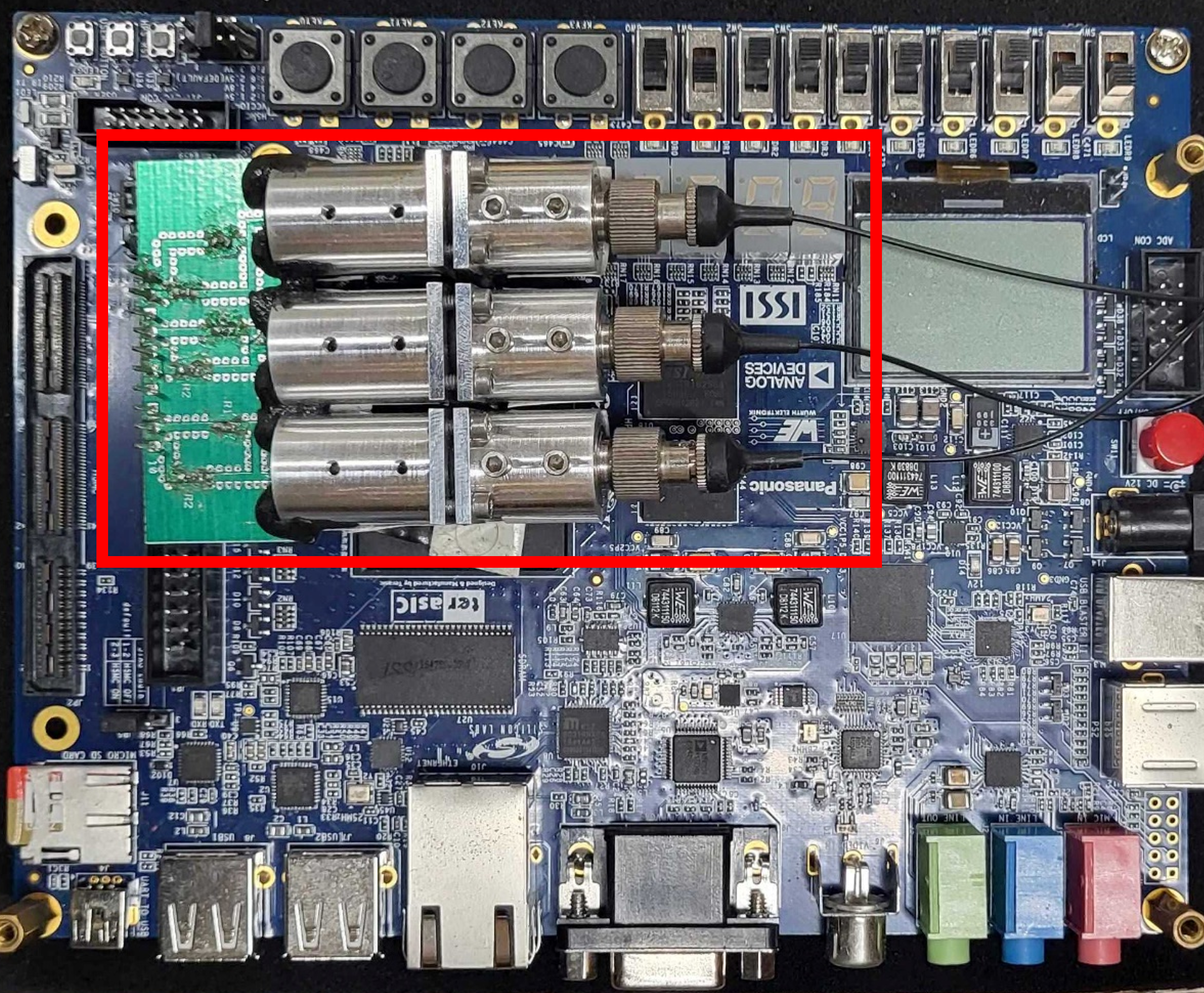}
\caption{Image showing the control system FPGA, the circuit board, and the RC-LEDs with their custom-built couplers.} 
\label{fig:FPGA_LED_coupling}
\end{figure}

Figure \ref{fig:FPGA_LED_coupling} shows the experiment apparatus with the FPGA, circuit board connecting the FPGA to the RC-LEDs, couplers, and SMFs.  The dimension of the custom package is 6.4 cm $\times$ 5.1 cm $\times$ 2.1 cm and weighs 0.12 kg. 

\subsubsection{Encoder Optical Setup} \label{subsub:TX_optical}

\begin{figure}[htbp]
\centering
\includegraphics[width=0.93\linewidth]{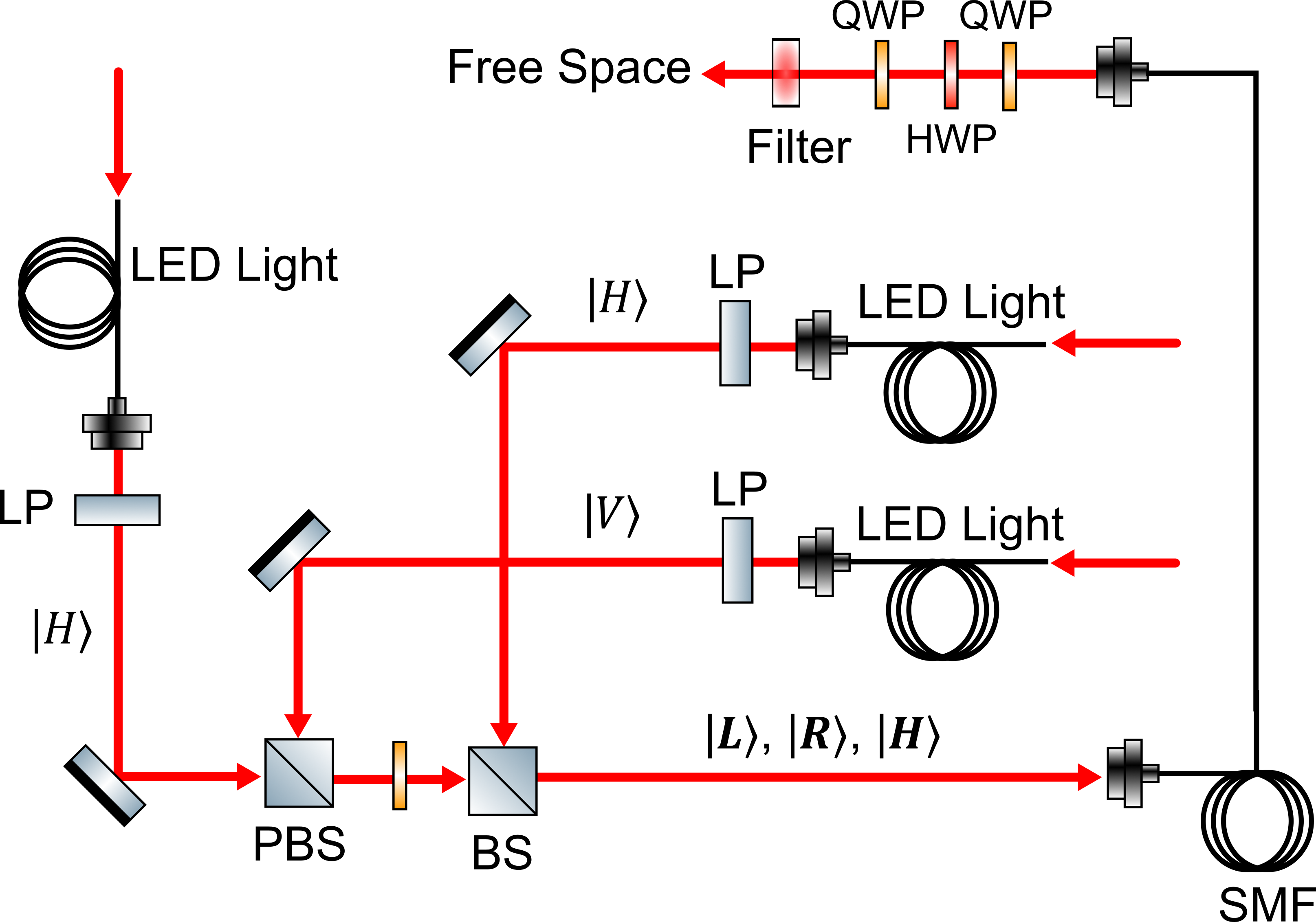}
\caption{Three-state QKD polarization encoder showing the layout of the optical components.} 
\label{fig:Optical_Setup_TX}
\end{figure}

The source light is brought onto the encoder optical bench using reflective collimators (Thorlabs RC02FC-P01). To encode the states, we use a series of linear polarizers (LP), polarizing (PBS) and non-polarizing (BS) beam splitters, and a quarter-wave plate (QWP) to create the three quantum photonic polarization states: $\ket{L}, \ket{R}$, and $\ket{H}$ as illustrated in Fig.~\ref{fig:Optical_Setup_TX}.

For the circularly polarized states, the light for each channel passes through LPs, transforming the unpolarized RC-LED light into the states $\ket{V}$ (middle right of the diagram) and $\ket{H}$ (left of the diagram). These states are combined with low loss using the PBS whose output passes through a QWP with the fast axis set at 45$^\circ$ from the horizontal, transforming $\ket{V}$ ($\ket{H}$) to $\ket{L}$ ($\ket{R}$). The third state $\ket{H}$ is created with an LP (top right of the diagram) and combined with the $\ket{L}$/$\ket{R}$ states using a BS with an efficiency of 50\%. All states are then coupled to a SMF, subsequently passing through a series of waveplates for state compensation (see Sec.~\ref{sub:indistinguishability}), and a spectral filter. 

The complete transmitter setup (FPGA, RCLEDs and coupler, and the encoder) measures roughly 1800 cm$^3$, weighs 2 Kg, and has a power consumption of 2.5W. 

\subsection{Indistinguishability} \label{sub:indistinguishability}

\begin{figure}[htbp]
\centering
\includegraphics[width=1.0\linewidth]{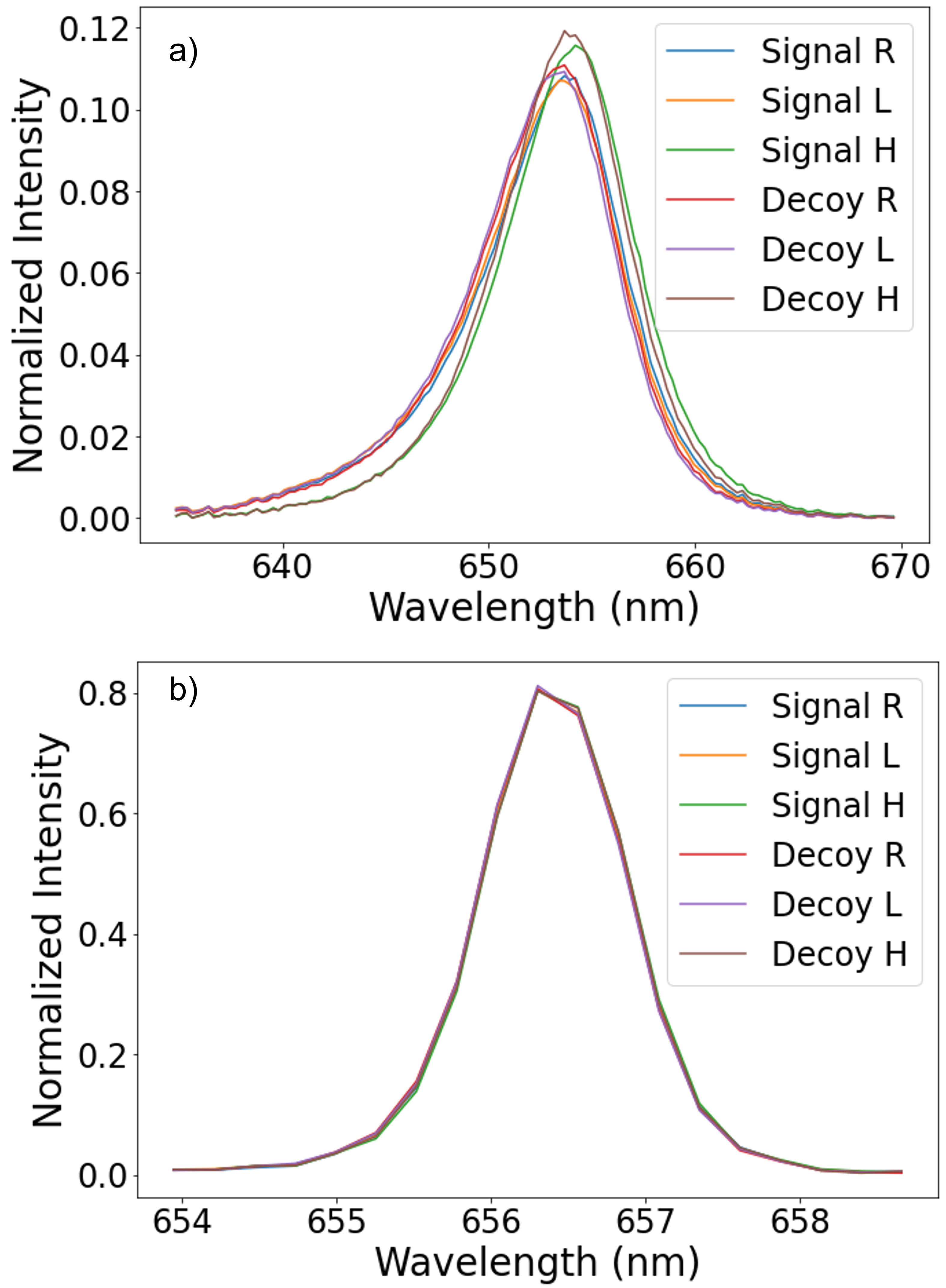}
\caption{RC-LED spectra for the different states. a) Unfiltered and b) filtered cases. The resolution of the spectrometer used in this measurement is 0.3 nm.}
\label{fig:10_11_22_spectra}
\end{figure}

We must ensure that photons in each of the three states are indistinguishable in the degrees-of-freedom other than polarization to protect the QKD system from side-channel attacks made directly on the transmitted quantum photonic states. For the spectral degree-of-freedom, we hand-select three RC-LEDs with the most similar spectra from a collection of 23 devices.  As seen in Fig.~\ref{fig:10_11_22_spectra}a, the full-width-at-half-maximum of the spectra are $\sim$7 nm and the emission peaks are within $\sim$2 nm.  Also seen is a spectral shift for the same RC-LED when driven by a lower current to create the decoy state.  Using the procedure given in Appendix \ref{app:quantifying_indistinguishability}, we find that the fraction mutual information between Eve and Bob for these states is equal to $1.77\times10^{-4}$ with a $1.2\times10^{-6}$ measurement bias and an uncertainty of $1.6\times10^{-5}$.

To increase the indistinguishability, we pass the combined states through a 1.2-nm-wide spectral filter (see Fig.~\ref{fig:Optical_Setup_TX}) with a center wavelength of 656.3 nm (Andover 656FS02-12.5), producing the spectra for each state shown in Fig.~\ref{fig:10_11_22_spectra}b. The fractional mutual information decreases by about an order-of-magnitude to $2.4\times10^{-5}$ with a $2.4\times10^{-5}$ measurement bias and uncertainty of $1.2\times10^{-5}$. 

We next discuss the temporal indistinguishability of the quantum photonic states. Temporal variations arise from multiple sources in our system, including differences in electrical and optical paths, and the change in the RC-LED rise time with current apparent when comparing the signal and decoy state optical waveforms. As described in Sec.~\ref{sub:FPGA_pulse_generation}, we adjust the delay of the PPL outputs to align and shape the waveforms. 

\begin{figure}[htbp]
\centering
\includegraphics[width=1.0\linewidth]{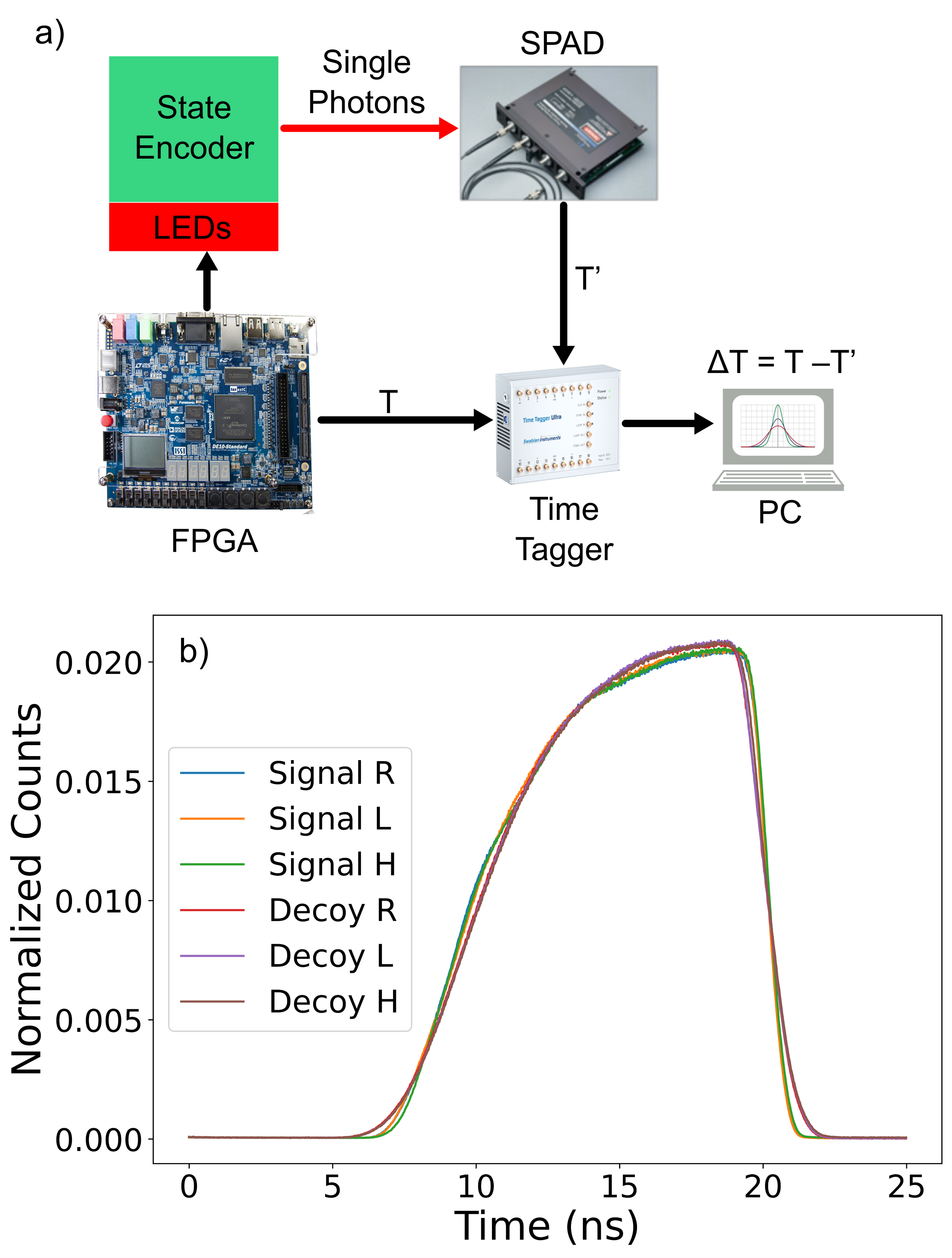}
\caption{a) Experimental setup for measuring the temporal waveform for each of the quantum photonic states. We generate a histogram with a bin width of 20 ps. b) Temporal profile for both signal and decoy states.}
\label{fig:Temporal_setup_2}
\end{figure}

We measure the optical waveform of each state by passing them, one at a time, through the entire electrical and optical sub-systems and measure the resulting wavepacket with a SPAD detector (SAP-500) with a jitter of $\sim$ 170 ps. The resulting electrical pulse time-of-arrival $T'$ is measured with a multi-hit time tagger (Swabian TimeTagger Ultra, with a jitter of 8 ps).  We further attenuate the light to a mean photon number of $\sim$ 0.1 to prevent waveform distortion. We also send the electrical pulse that drives the RC-LED to the time-tagger with arrival time denoted by $T$.  The waveform is created by histogramming the number of counts for the time difference $\Delta T = T - T'$. Typically, we collect $\sim 10^8$ events for each waveform.  The experimental setup used for the temporal compensation is shown in Fig.~\ref{fig:Temporal_setup_2}a. 

We use the following procedure to adjust the waveforms.  One state, selected at random, is the reference to which all other states are compared. For the other waveforms, we compare their profiles to the reference and adjust the shift and width using the corresponding PLLs in increments of 78 ps.  We typically iterate this procedure $\sim$ 5 times.

A typical final configuration is shown in Fig.~\ref{fig:Temporal_setup_2}b. The decoy states have slightly different rise and fall times compared to the signal states, which is the leading contribution to the remaining distinguishability of the states. We quantify the indistinguishability of the sources using the fractional mutual information leaked to an eavesdropper that measures the spectral or temporal degree-of-freedom.  The derivation in the Appendix shows how the mutual information can be calculated from the spectral or temporal probability distribution functions for each state and the sending probability of each state.  For this data, and modulo some assumptions of possible unresolved structure in the spectrum and waveform for the LEDs (see Appendix), we obtain the fractional mutual information between Eve and Bob of $4.31\times10^{-5}$ with a measurement bias of $3.18\times10^{-6}$ and a $7.99\times10^{-7}$ uncertainty. 

For the spatial degree-of-freedom, we couple all three optical paths into the same 0.3-m-long SMF to guarantee spatial indistinguishability \cite{Wallner}. We use two variable focal length collimators (Thorlabs CFC11A-A) to couple in and out of the SMF. This allows us to control the outgoing beam size of the QKD signal, which is important for experiments performed at different distances.

Passing light through an SMF, however, can transform the incoming polarization states with an unknown unitary transformation, lowering the QKD system state fidelity. We use a combination of two quarter-wave plates and one half-wave plate to compensate for this distortion (see Fig.~\ref{fig:Optical_Setup_TX}). These waveplates are tuned before every experiment because changes in the fiber positioning modify the unitary transformation. 

\subsection{QKD Receiver and Data Processing} \label{sub:Receiver}

\begin{figure}[htbp]
\centering
\includegraphics[width=0.9\linewidth]{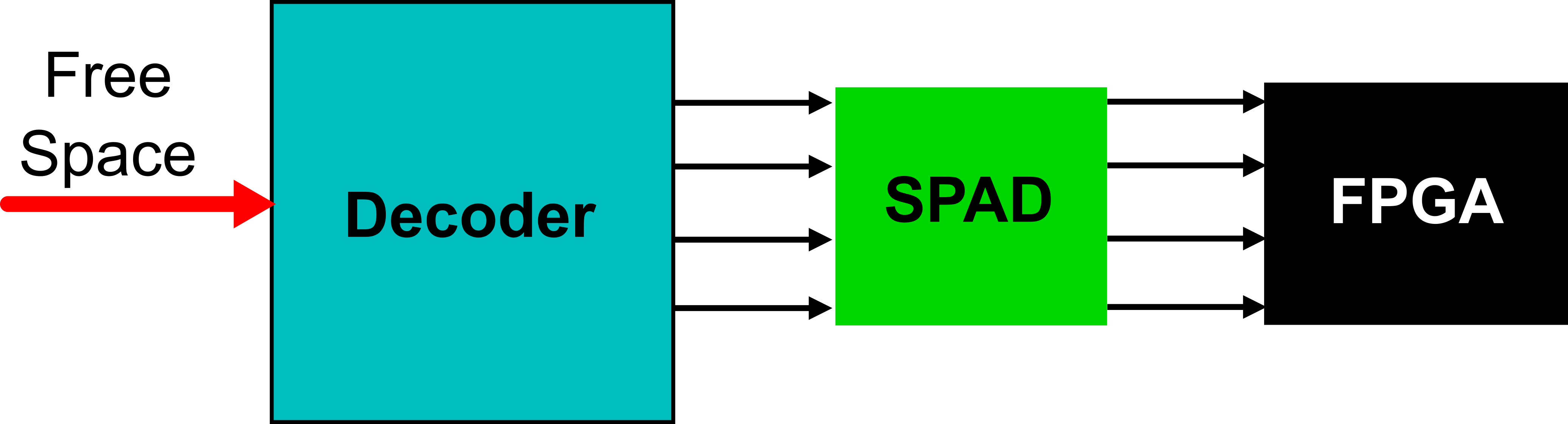}
\caption{High-level diagram showing the experimental setup for Bob's QKD receiver.} 
\label{fig:high_level_RX}
\end{figure}

Like the transmitter, the receiver has three key components illustrated in Fig.~\ref{fig:high_level_RX}.  Received wavepackets are decoded using a free-space optical bench, each decoded state is passed to a SPAD using a multi-mode fiber, and an FPGA is used at the control system for measuring the event time of arrival and processing the data.  Below, we describe each of these sub-systems. 

\subsubsection{State Decoding} \label{sub:state_decoding}

\begin{figure}[htbp]
\centering
\includegraphics[width=0.8\linewidth]{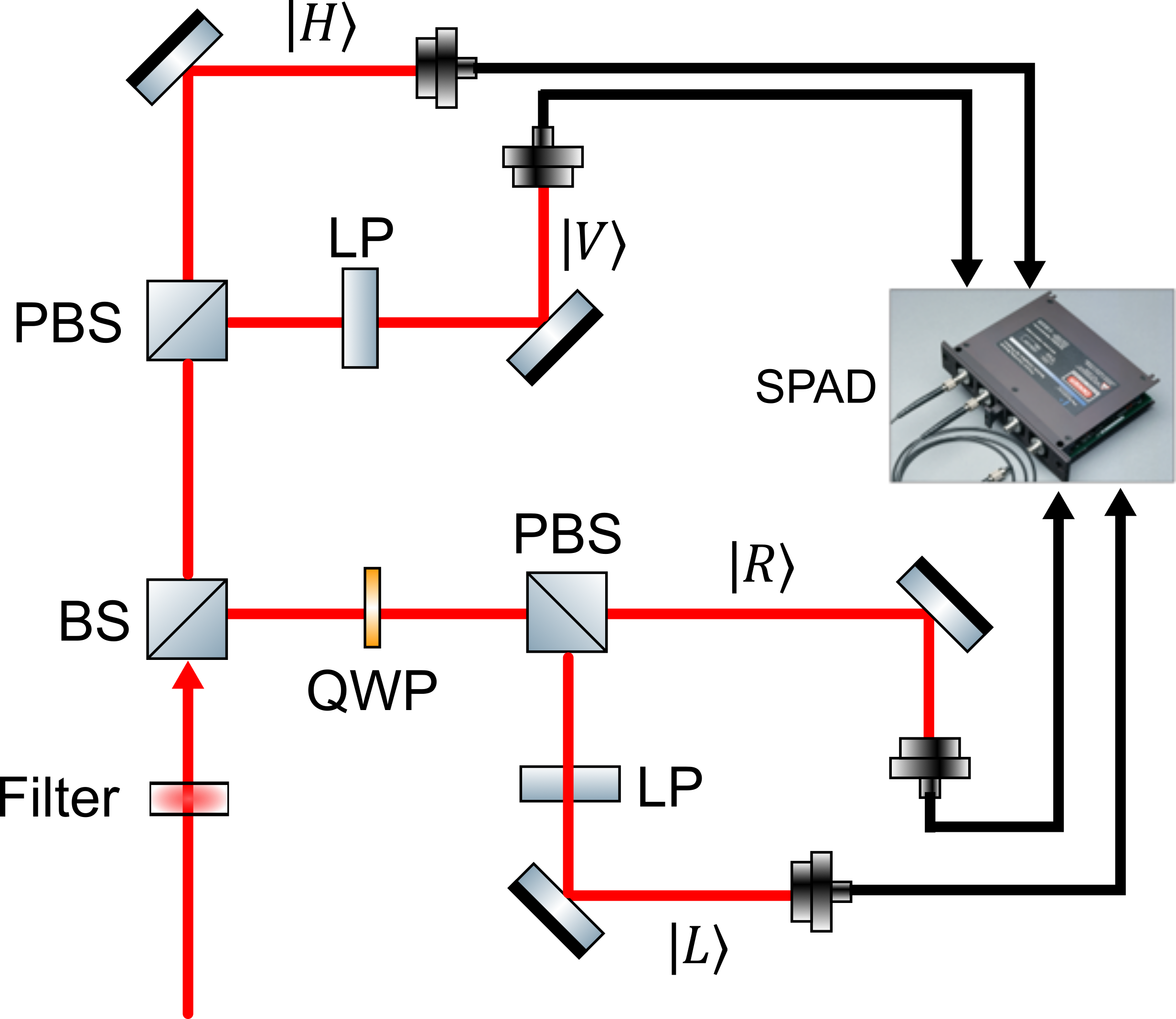}
\caption{Optical setup used to decode the states generated by Alice.}
\label{fig:Optical_Setup_RX}
\end{figure}

The optical decoder, shown in Fig.~\ref{fig:Optical_Setup_RX}, sorts and measures the four states of polarization: $\ket{L}, \ket{R}, \ket{H}$, and $\ket{V}$. The incoming free-space quantum photonic states pass through a 1.2-nm-wide spectral filter (Andover 656FS02-12.5), which matches the transmitter filter and minimizes noise from ambient light. Additionally, the receiver is covered by a light-proofing black plastic 3D-printed box. We use one-inch optics throughout the decoder to increase the signal collection efficiency.

The polarizations are sorted using a series of NPBS, PBS, and a QWP. First, the BS randomly selects a basis to be measured (either L/R or H/V) with equal probability. On the L/R arm, we add a QWP to transform the circularly polarized light to the states $\ket{V}$ and $\ket{H}$. The PBS sorts these states and directs them to multi-mode fibers (Thorlabs FT200UMT, 200-$\mu$m-diameter core, 0.39 numerical aperture), which are connected to SPAD detectors.  On the H/V arm, the states are sorted by a PBS and passed to the detectors.

The receiver setup (decoder, detectors, and FPGA) measures roughly 2728 cm$^3$, weighs 2.7 Kg, and has a power consumption of 5.9 W. 

\subsubsection{Data Processing} \label{subsub:RX_data_processing}

The electrical pulses generated by the SPADs are passed to the FPGA using 50-$\Omega$-impedance coaxial cables. The voltage pulses produced by the SPADs have an amplitude of 5 V, whereas the FPGA accepts 3.3 V.  We fabricated a custom resistive splitter on a small circuit board, which is inserted into the GPIO connector that reduces the voltage and impedance-matches the cable. 

The received electrical pulses are time-tagged on the FPGA using a 30-bit counter incremented at every positive edge of a PLL-derived 100 MHz clock.  Thus, the counter rolls over every 10.7 s. At the arrival of a detector signal, the FPGA saves the current 30-bit counter value to the FPGA's on-chip RAM.

The time-tagged data are temporarily stored in four independent 128-kB-deep banks of on-chip FPGA RAM. When any bank acquires 64 KB of data, the FPGA signals the hard processor to move that data into a 600 MB buffer located on the hard processor's 1 GB synchronous dynamic RAM (SDRAM). When the 600 MB buffer is full, the hard processor saves the data to the SD card.  All events are time-tagged in parallel so that simultaneous events, such as from a multi-photon wavepacket whose photons are sorted to different detectors, are recorded. Recording simultaneous events is an important input to the system security analysis. 

A similar procedure is adopted on the transmitter. We generate a new random state at a rate of 12.5 MHz. Every generated state is encoded using 3 bits and appended to a 30-bit word, which are temporarily stored in a 128-kB-deep memory bank on the FPGA. When half of the bank is full (typically every 13.1 ms), the FPGA signals the hard processor to move the 64-kB-long data segment to a 32-MB-deep buffer on its SDRAM. When the buffer is full (typically every 6.71 s), all data generation and transmission are halted for 0.5 s to allow enough time for the hard processor to save the buffer to the SD card. 

While the crystal oscillators on each FPGA demonstration board have high stability over a typical QKD session, their frequencies are slightly different and their counters do not begin simultaneously. Thus, the transmitted and received data are not temporally aligned. Synchronizing Alice and Bob's data is essential for sifting, where Alice and Bob publicly discuss their basis choices and keep only the data when Bob receives a detector click and when they are prepared and measured in the same basis. 

We synchronize the data using a post-processing algorithm we developed that is based on a Bayesian method \cite{Roddy} that recovers the clock offset between the transmitter and receiver using only the published basis and decoy state choices. Thus, it does not reveal any information to Eve because the basis and decoy state choices are already revealed over the public channel for the sifting stage of the analysis. We discard any data where the synchronization confidence is below 95\%. 

Also during post-processing, we estimate the quantum bit-error rate (QBER) calculated for each channel ($\ket{R}$, $\ket{L}$, and $\ket{H}$) and given by
\begin{eqnarray}
    R_{QBER} &=& \frac{N_L^R}{N_L^R + N_R^R}, \\
    L_{QBER} &=& \frac{N_R^L}{N_R^L + N_L^L}, \\
    H_{QBER} &=& \frac{N_V^H}{N_V^H + N_H^H},
\end{eqnarray}
where $N_X^Y$ is the number of events in detector $X$ given that $Y$ wavepackets were sent. 

\section{QKD Results} \label{sec:results}

\subsection{Tabletop QKD} \label{sub:results_tabletop}

We test our apparatus in the lab in a tabletop configuration. We first maximize RC-LED light coupled into the SMF from Sec.~\ref{subsub:LED_optical}. We then align the optical components of Alice's side and maximize the light coupled into the spatial filter fiber from Sec.~\ref{sub:indistinguishability}. We follow this by performing temporal compensation, as discussed in Sec.~\ref{sub:indistinguishability}. Bob's setup is aligned by back-propagating laser beams through each of Bob's multi-mode fibers and adjusting the optics to simultaneously overlap the beams close to the bench and at a distance of $\sim$10~m. 

Once both Alice and Bob's setups are aligned, we separate the transmitter and receiver by a distance of $\sim$1.5 m on an optical table and align them to each other using auxiliary mirrors. We then compensate the polarization states emitted by the transmitter, as discussed in Sec.~\ref{sub:indistinguishability}. 

\begin{figure}[htbp]
\centering
\includegraphics[width=1.0\linewidth]{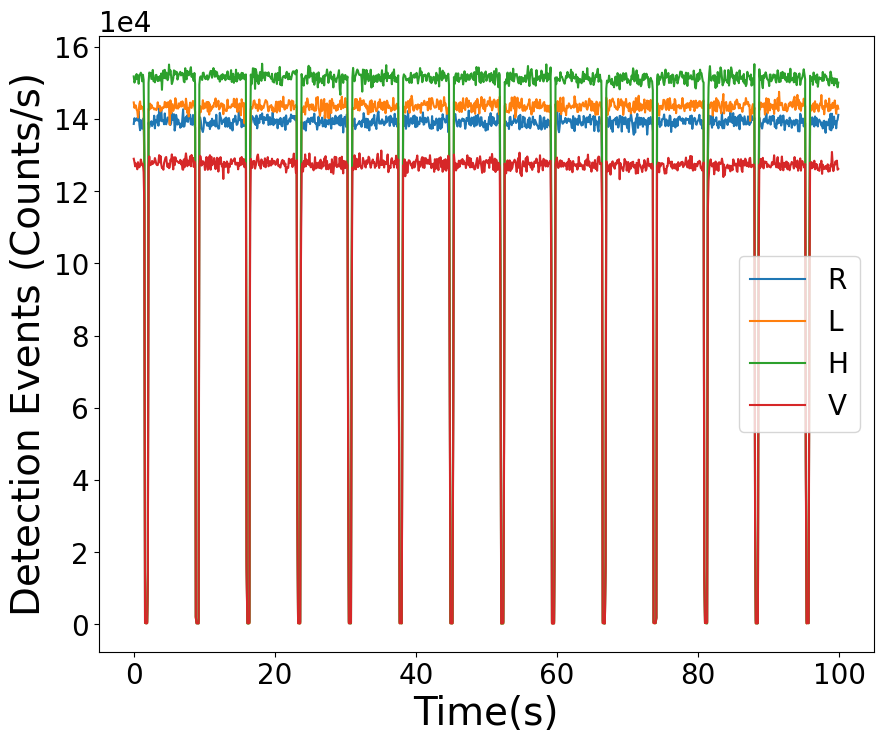}
\caption{Typical count rates on Bob's detectors for a tabletop QKD experiment.}
\label{fig:tabletop}
\end{figure}

Finally, we collect data for about 100 s, which typically generates 5$\times$10$^7$ events on Bob's side. Fig.~\ref{fig:tabletop} shows the typical count rate for each state during a QKD session. Notably, the periodic drops in counts are intentional and occur when the transmitter FPGA halts all operations for 0.5 s to save data to the SD card as discussed in Sec.~\ref{subsub:RX_data_processing}. 

We obtain QBERs of 1.93\% in the $\ket{R}$ channel, 1.50\% in the $\ket{L}$ channel, and 2.05\% in the $\ket{H}$ channel.  Here, we estimate the errors using the entire dataset shown in Fig.~\ref{fig:tabletop}, but the results will be similar using a shorter dataset given the high stability of the count rates in this tabletop setup. All QBERs are far below the 11\% limit for which QKD protocols based on BB84 will not yield a secure private key \cite{11percent, QI_Lectures, QBER_threshold}. Additionally, we obtain raw key rates of up to 532 Kbits/s. 

\section{Conclusions} \label{sec:conclusion}

We describe the design considerations for a QKD system suitable for mobile applications and use it for a laboratory QKD session.  This system has been used for drone-to-drone, drone-to-car, and car-to-car QKD, as reported elsewhere \cite{SPIE_1, SPIE_2}. 

In the future, we plan several improvements to our system. Regarding SWaP, one improvement is to develop a compact, potentially monolithic encoder and decoder \cite{Hand_Held} that will reduce the SWaP and improve the alignment stability. Similarly, we found the RC-LED - to - SMF coupler misaligned over days due to the relaxation of the o-ring in the coupler.  A new coupler design using pre-aligned optics may be beneficial. 

Another bottleneck in data processing is the off-line clock synchronization. We are currently developing a method to synchronize the clocks on each FPGA to Uniform Coordinated Time using a global positioning system (GPS) timing receiver.  

\section{Acknowledgments}

We gratefully acknowledge the financial support of NASA  (\#80NSSC20K0629) and the U.S. Office of Naval Research Multidisciplinary University Research Initiative program on `Wavelength-Agile Quantum Key Distribution in a Marine Environment' (\#N00014-13-1-0627). DSR and DJG gratefully acknowledge the financial support of the U.S. Air Force Research Laboratory project `Center for Enabling Cyber Defense in Analog and Mixed Signal Domain' (\#FA8650-19-1-1741). RDC and DJG gratefully acknowledge the financial support of the U.S. Air Force Research Laboratory and the Southwestern Council for Higher Education (\#FA8650-19-2-9300).  SI and PGK gratefully acknowledge the financial support of the U.S. Army Research Laboratory project `Mobile-Platform Quantum Communication' (\#W911NF2420097), the KRISS Korea project `Development of Core Technologies for Entanglement-Based Multi-Party Hybrid Quantum Networks' (NRF \#112168), and the U.S. Department of Energy project `Advanced Quantum Networks for Science Discovery (DOE Fermi 712869). The U.S. and Korean governments are authorized to reproduce and distribute reprints for governmental purposes notwithstanding any copyright notation thereon. The views and conclusions contained herein are those of the authors and should not be interpreted as necessarily representing the official policies or endorsements, either expressed or implied, of the funding agencies listed above or the U.S. or Korean governments.

\section{Code Availability}
All code for this project is available \cite{code}.  The Verilog code developed for this project targets the DE10-Standard FPGA board using the Quartus Prime Lite 18.1 environment.  Additionally, the code to analyze the spectra, temporal profiles, tabletop, and indistinguishability data is developed in Python.

\section{Appendix} \label{sec:appendix}
\subsection{Quantifying Indistinguishability} \label{app:quantifying_indistinguishability}
In our system, we use pulses generated by RC-LEDs.  These pulses are not transform-limited (\textit{i.e.}, their spectral bandwidth is $\sim$1,000$\times$ broader than the inverse pulse width).  Hence, there is, in principle, undetected distinguishing information in the wavepackets.  For example, there could be a unique temporal pattern in the pulses that is faster than the resolution of the SPADs, a fine spectral pattern below the spectrometer resolution, or a frequency or phase modulation, etc. unique to each RC-LED.  An eavesdropper with a faster response-time detector, a higher resolution spectrometer, or a more general pulse analysis method, could use these patterns to distinguish between the three LEDs.  

Here, we assume that thermal states describe the RC-LEDs without such identifying fine structure in the spectral-temporal modes, and hence the as-measured states can be used to quantify the indistinguishability.  This assumption is physically reasonable because no obvious mechanisms would lead to such a fine structure.  For example, while a frequency chirp in the LED spectrum could give such structure, it would need to be a different chirp for each LED to create unwanted distinguishing information, whereas the LEDs are all of the same construction and the photons pass through similar optics before leaving the transmitter.

Using our as-measured spectral-temporal measurements, we estimate the indistinguishability of Alice's states using the mutual information to calculate the fraction of the final sifted key potentially shared between Bob and Eve due to the spectral and temporal differences.  We denote $E$ as the set of Eve's side channel measurements, $B$ as Bob's sifted bits, and $S$ as the statement that the channel access attempt resulted in a sifted bit.  The mutual information between Eve's measurements and Bob's sifted key is given by
\begin{equation}\label{eq:mutualinfo}
I(B;E|S) = H(B|S) + H(E|S) - H(B,E|S),
\end{equation}
where $H$ is the Shannon entropy
\begin{equation}
H(X) := -\displaystyle \sum_{x \in X} p(x)\log p(x).
\end{equation}

Thus,

\begin{eqnarray}
I(B;E|S) = -\displaystyle \sum_{k} p(B_k|S)\log p(B_k|S) \\ \nonumber
-\displaystyle \sum_{j} p(E_j|S)\log p(E_j|S) \\ \nonumber
+ \displaystyle \sum_{k,j} p(B_k,E_j|S)\log p(B_k,E_j|S). \label{eqn:mutual_expanded}
\end{eqnarray}

To compute the entropies, we must find $p(B_k|S)$, $p(E_j|S)$, and $p(B_k,E_j|S)$ in terms of known probabilities.

The probability $p(E_j|A_i)$ is the conditional probability of the eavesdropper measuring a state $j$ (such as a particular wavelength in the spectral measurement, or a particular arrival time in the temporal measurement), is given by the distributions in Figs. \ref{fig:10_11_22_spectra}b and \ref{fig:Temporal_setup_2}b.  The probability $p(A_i)$ is the sending probability for each state, $p(S|A_i)$, which is the likelihood of a particular state resulting in a sifted bit, and $p(B_k|A_i,S)$, which is the probability of the sifted bit result recorded by Bob for a given input state.  These latter three can be calculated from the sending state fraction, the expected receiver sorting, and the estimated channel loss.

The conditional probability $p(B_k|S)$ can be expanded using marginalization (\textit{i.e.}, summing over some variables), giving
\begin{equation}
p(B_k|S) = -\displaystyle \sum_{i} p(B_k|A_i,S)p(A_i|S),
\end{equation}
where $A$ is the set of states Alice can send.  Applying Bayes' Theorem, we find
\begin{equation}
p(B_k|S) = -\displaystyle \sum_{i} p(B_k|A_i,S)\dfrac{p(S|A_i)p(A_i)}{p(S)}.
\end{equation}
Applying marginalization again, we find
\begin{equation}
p(B_k|S) = -\dfrac{\displaystyle \sum_{i} p(B_k|A_i,S)p(S|A_i)p(A_i)}{\displaystyle \sum_{\ell}p(S|A_\ell)p(A_\ell)},
\end{equation}
with all terms now being known probabilities.

The conditional probability $p(E|S)$ can be found using a similar procedure, resulting in the relation
\begin{equation}
p(E_j|S) = -\dfrac{\displaystyle \sum_{i} p(E_j|A_i,S)p(S|A_i)p(A_i)}{\displaystyle \sum_{\ell}p(S|A_\ell)p(A_\ell)},
\end{equation}
which can be simplified using the knowledge that $p(E_j|A_i,S) = p(E_j|A_i)$ because the distribution of Eve's measurements is fully determined by knowledge of Alice's state, and hence $S$ is irrelevant. Using these relations, we find
\begin{equation}
p(E_j|S) = -\dfrac{\displaystyle \sum_{i} p(E_j|A_i)p(S|A_i)p(A_i)}{\displaystyle \sum_{\ell}p(S|A_\ell)p(A_\ell)}.
\end{equation}
We can rewrite
\begin{eqnarray}
p(B_k,E_j|S) = p(B_k|E_j,S)p(E_j|S), \\
p(B_k,E_j|S) = \displaystyle \sum_{i} p(B_k|A_i,E_j,S)p(A_i|E_j,S)p(E_j|S), \\
p(B_k|A_i,E_j,S) = p(B_k|A_i,S),
\end{eqnarray}
because Bob's measurement outcome is fully determined by Alice's input state and the knowledge that it results in a sifted bit.  Because Eve's measurement is a side channel measurement in this derivation, it has no impact on Bob's measurement in the chosen degree of freedom.

Finally, we obtain
\begin{eqnarray}
p(B_k,E_j|S) = \displaystyle \sum_{i} p(B_k|A_i,S)p(A_i|E_j,S)p(E_j|S), \\
p(B_k,E_j|S) = \displaystyle \sum_{i} p(B_k|A_i,S)\dfrac{p(E_j|A_i,S)p(A_i|S)}{p(E_j|S)}p(E_j|S), \\
p(B_k,E_j|S) = \displaystyle \sum_{i} p(B_k|A_i,S)p(E_j|A_i,S)\dfrac{p(S|A_i)p(A_i)}{p(S)}, \\
p(B_k,E_j|S) = \dfrac{\displaystyle \sum_{i} p(B_k|A_i,S)p(E_j|A_i,S)p(S|A_i)p(A_i)}{\displaystyle \sum_{\ell}p(S|A_\ell)p(A_\ell)}.
\end{eqnarray}
We have now written all relevant quantities in terms of known probabilities that can be inserted into Eq.\ref{eqn:mutual_expanded} to find the mutual information Eve gains from the side-channel measurements.

However, we must also account for the systematic bias in our mutual information calculation.  Because measurement noise inherently makes a variable more random, it creates a systematic bias for higher entropies.  Roulston \cite{ROULSTON} shows that the observed entropy $H_{obs}$ can be written as
\begin{equation}\label{eq:entropy_bias}
\langle H_{obs} \rangle= H_{\infty} + \displaystyle \sum_{i}^B \dfrac{\langle \epsilon_i^2 \rangle p_i}{2},
\end{equation}
where $H_{\infty}$ is the `true' entropy of the system, $B$ is the number of states, $\epsilon_i$ is the fractional measurement error in the state probability $p_i$, with the measured probability $q_i$ such that 
\begin{equation}
\epsilon_i = \dfrac{q_i-p_i}{p_i}.
\end{equation}

Thus, Eq.~\ref{eq:entropy_bias} can be rewritten as
\begin{equation}
\langle H_{obs} \rangle= H_{\infty} + \displaystyle \sum_{i}^B \dfrac{\langle (q_i-p_i)^2 \rangle}{2 p_i},
\end{equation}
or
\begin{equation}\label{eq:entropy_bias_known}
\langle H_{obs} \rangle= H_{\infty} + \displaystyle \sum_{i}^B \dfrac{\sigma_i^2}{2 q_i},
\end{equation}
using the definition of the variance and approximating the denominator as $2 q_i$, and $\sigma_i$ is the error in the measurement of $p_i$.  We now have the entropy bias in terms of known quantities, and we can add the biases of the constituent entropies to find the bias for the mutual information as a whole.

To quantify the uncertainty of the mutual information, we used two approaches.  The first is the standard propagation of errors technique using derivatives given by
\begin{equation}
\sigma_f = \sqrt{\displaystyle \sum_{i}\bigg(\pdv{f}{x_i}\bigg)^2 \sigma_{x_i}^2}. \nonumber
\end{equation}
That is, for function $f$, the uncertainty $\sigma_f$ in terms of the uncertainties of the constituent variable $x_i$ and its derivatives.

However, this formula is an approximation that assumes $\partial f/\partial x_i$ is approximately constant in the neighborhood of $\sigma_{x_i}$.  While this approximation is accurate for evaluating the temporal degree-of-freedom (for which our measurement has a smaller error), it is inaccurate for the spectral degree-of-freedom.

For the spectral data, we use a Monte-Carlo sampling approach.  By randomly sampling the constituent variables from their distributions and evaluating the mutual information function many times, we create a histogram representing the probability distribution of the mutual information.  The standard deviation of this distribution is the uncertainty of the mutual information measurement.
% Bibliography
\bibliography{MobileQKD}
\end{document}